# Aligning Software-related Strategies in Multi-Organizational Settings

*Martin Kowalczyk[A], Jürgen Münch[A], Masafumi Katahira[B],*

*Tatsuya Kaneko[B], Yuko Miyamoto[B], Yumi Koishi[B]*

[A]Fraunhofer IESE, [B]Japan Aerospace Exploration Agency (JAXA)

{Martin.Kowalczyk, Juergen.Muench}@iese.fraunhofer.de

{Katahira.Masafumi, Kaneko.Tatsuya, Miyamoto.Yuko, Koishi.Yumi}@jaxa.jp

*Abstract:*

*Aligning the activities of an organization with its business goals is a challenging task that is critical for success. Alignment in a multi-organizational setting requires the integration of different internal or external organizational units. The anticipated benefits of multi-organizational alignment consist of clarified contributions and increased transparency of the involved organizational units. The GQM+Strategies approach provides mechanisms for explicitly linking goals and strategies within an organization and is based on goal-oriented measurement. This paper presents the process and first-hand experience of applying GQM+Strategies in a multi-organizational setting from the aerospace industry. Additionally, the resulting GQM+Strategies grid is sketched and selected parts are discussed. Finally, the results are reflected on and an overview of future work is given.*

## 1 Introduction

Aligning organizational activities with top-level business goals is highly relevant, particularly in difficult economic situations. Aligned organizations are able to act with higher effectiveness and efficiency and thus can achieve competitive advantages [2]. The GQM+Strategies[1] approach [3] helps to harmonize and communicate goals and strategies across multiple levels of an organization and therefore makes it possible to align goals and strategies across the levels of an organization's hierarchy. Additionally, the approach supports monitoring the success or failure of strategies and the fulfillment of associated business goals. The application of the GQM+Strategies approach generates a model of linked goals and strategies and defines a strategic measurement system. The resulting construct is called the GQM⁺Strategies grid [4, 5] and is the central element of the approach for managing organizational goals and strategies.

---

1 GQM+Strategies is registered trademark No. 302008021763 at the German Patent and Trade Mark Office; international registration number IR992843.

In the context of the GQM⁺Strategies application discussed in this paper, we address the following challenges: (1) An internal organizational unit wanted to explicitly highlight its contribution to top-level organizational goals and make sure that these contributions are aligned. Internal organizational units often have to clarify and advocate their contributions towards top-level business goals. In particular when it comes to budget allocation, this ability becomes crucial and measurement-based alignment can become a differentiator. (2) Another objective was to align and make transparent the measurement needs in the context of distributed projects, explicitly including external suppliers. Increasing the transparency of distributed projects through measurement was seen as a key component for the improvement of overall project success. Defining measurement systems in the context of distributed collaborations represents a challenge, as different understandings and motivations with respect to measurement have to be aligned.

This context of the presented GQM⁺Strategies application addresses a multi-organizational setting as it includes internal units or external organizations. Business alignment in a multi-organizational setting requires the integration of these different organizations, and we will show how the GQM⁺Strategies approach was used for this purpose.

Section 2 gives an overview of related work and basic concepts. Section 3 presents the systematic deployment process for GQM⁺Strategies and describes its specific deployment at the Japan Aerospace Exploration Agency (JAXA). Section 4 discusses modeling aspects and the resulting GQM⁺Strategies grid that was developed at JAXA. Section 5 presents lessons learned and improvement potentials. Finally, section 6 concludes this paper and provides an outlook on future work in the context of GQM⁺Strategies.

## 2      Related Work and Basic Concepts

### 2.1      Related Work

In the area of software measurement, several approaches have been developed in the past. Goal-oriented approaches such as the GQM approach [1] offer the advantage that they guide data selection and analysis in a systematic way. These approaches are usually employed on the project level. In addition, top-level approaches for organizational performance measurement exist, of which the most prominent one is the Balanced Scorecard (BSC) [7]. The BSC provides a high-level framework for defining measures for different organizational perspectives. One weakness of this approach is the explicit linkage of goals, strategies, and measures to the operational level. The GQM⁺Strategies approach [3, 4] was developed to support the definition and explicit linkage of goals, strategies, and associated measurement across multiple organizational levels. Thus, the approach does not only help to harmonize goals and strategies across multiple levels of an organization but also provides a strategic measurement system for the effective control of goal achievement and the success or failure of strategies [5].

The focus of the approach and of this paper lies on the definition of measurement models and their alignment and not so much on data collection, analysis, and visualization, which is often the focus of Business Intelligence approaches.

### 2.2 Basic GQM⁺Strategies Concepts

Giving a short overview of the conceptual GQM⁺Strategies model will facilitate understanding of the actual implementation of GQM⁺Strategies at JAXA, as so far only few application examples are available.

The GQM⁺Strategies grid represents the central component of the GQM⁺Strategies approach and consists of two basic elements; Goal+Strategies elements and GQM graphs (see Figure 1). The former specify goals and strategies across different organizational levels and capture associated context and assumptions. The latter provide measurement models for controlling the associated Goal⁺Strategies elements. Goal⁺Strategies elements can be refined for the different levels of an organization's hierarchy and such a set of Goal⁺Strategies elements and GQM graphs constitutes a Goal⁺Strategies grid (see also [5]).

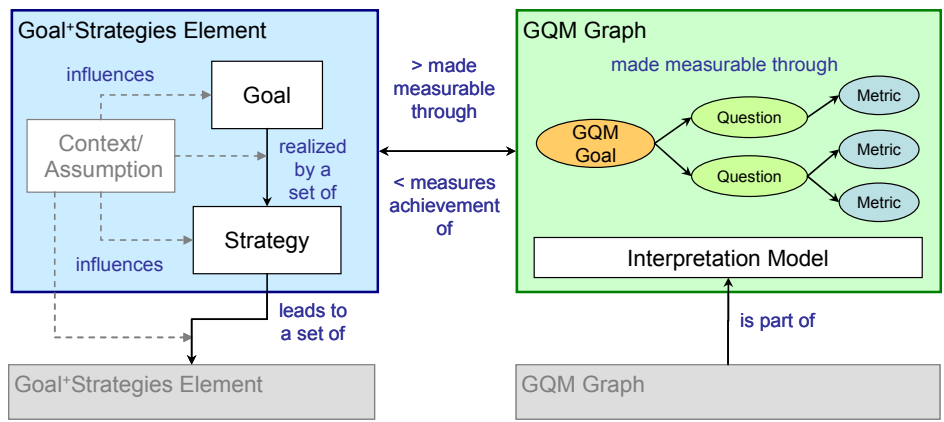

**Figure 1:** Goal⁺Strategies element (left) and GQM graph (right) (adapted from [5])

### 3 Applying GQM⁺Strategies

Introducing a new methodology to industrial practice can be difficult when the application can not be performed efficiently, as resources for experimenting with new approaches are often scarce. A systematic introduction process accompanied by adequate tool support is therefore beneficial. The following sections provide an overview of the systematic deployment process for GQM⁺Strategies and then present in detail the actual performance of the process in the context of JAXA.

### 3.1 The GQM⁺Strategies Process

The GQM⁺Strategies deployment process supports the introduction and application of strategic measurement systems with GQM⁺Strategies by structuring the implementation steps and by providing tools that enhance modeling, visualization, and communication. The GQM⁺Strategies process was created by measurement experts from the Fraunhofer Institute for Experimental Software Engineering (IESE) and the Fraunhofer Center for Experimental Software Engineering (CESE) on the conceptual basis of the Quality Improvement Paradigm (QIP) [6].

The process consists of seven activities, namely Initialize, Characterize, Set Goals, Choose Process, Execute Process, Analyze, and Package.

*Initialize* prepares the application of GQM⁺Strategies by assuring commitment and required resources as well as defining responsibilities and planning the further course of actions. *Characterize* defines the scope (i.e., the objectives and parts of the organization) for the application of GQM⁺Strategies. *Set Goals* represents the modeling activity within the process that defines and aligns goals and strategies within an organization, defines appropriate measurement with GQM, and captures the results within the GQM⁺Strategies grid. The grid derivation process has been described in detail in [5]. The GQM⁺Strategies process supplements the grid derivation with the elicitation of existing assets (i.e., goals, strategies, measures, etc.) and a gap analysis for identifying misalignments, ambiguities, or missing elements. The gap analysis supports the modeling of grids by identifying and reusing existing assets. The next activity is *Choose Process,* which makes measurement operational. The result of this activity is a measurement plan. During *Execute Process,* data collection and validation are performed. *Analyze* encompasses data analysis and visualization. Analysis will most likely make changes to some goals or strategies necessary. Therefore, *Package* provides the activities for capturing changes and adapting the grid where necessary, thus keeping the measurement system up to date.

### 3.2 Applying GQM⁺Strategies at JAXA

The focus of the application of GQM⁺Strategies at JAXA was on the first three activities (Initialize, Characterize, Set Goals) and thus on modeling the GQM⁺Strategies grid. Due to the distributed nature of our collaboration, we proceeded in four main steps during the application of GQM⁺Strategies. The first step was a preparative step that included activities from Initialize and Characterize, followed by iterations for the modeling (Set Goals) of the grid (see Figure 2). This section will discuss each step in more detail.

**Preparation**

The activity *Initialize* was performed in a joint workshop. Commitment, resources, and responsibilities were clarified beforehand. As JAXA already has experience in applying GQM measurement, a good foundation for the application of GQM⁺Strategies existed. The initial workshop was performed with a small group consisting of one manager from middle management and one senior engineer. Due to a tight schedule in this distributed collaboration, we started with a GQM⁺Strategies mini-tutorial and provided information about the approach in a very condensed way in order to create a common understanding between the participants.

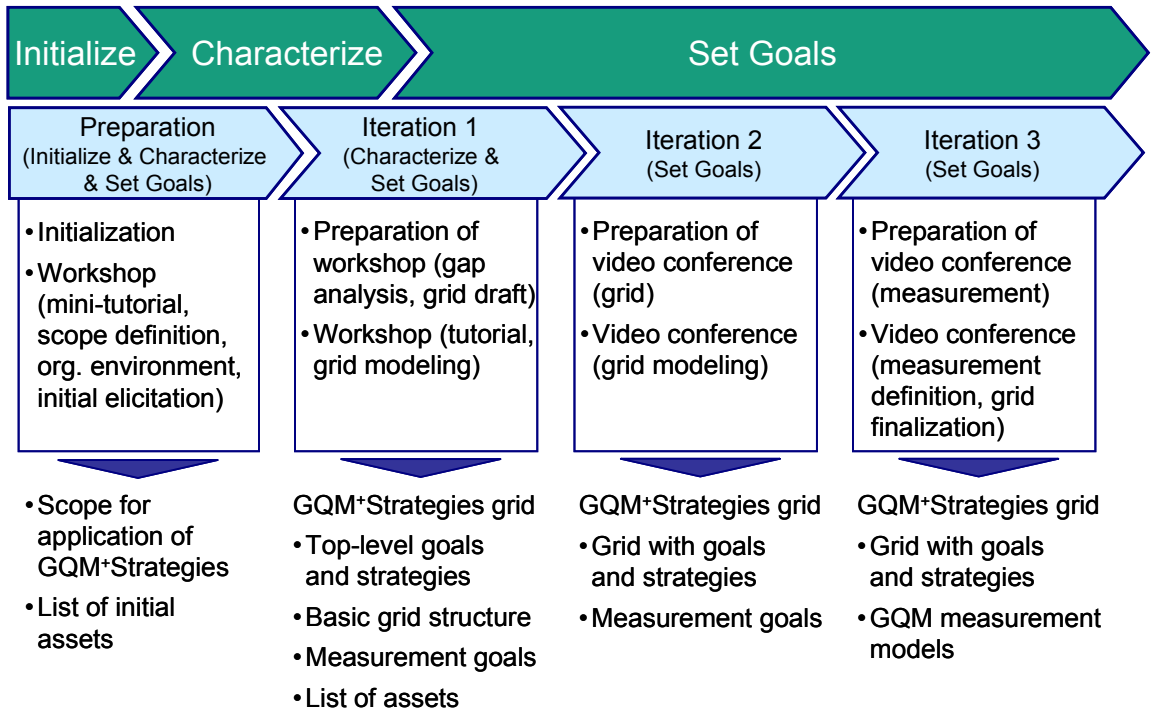

**Figure 2:** Overview of the application of GQM$^+$Strategies at JAXA

The measurement and GQM experience of all participants facilitated a quick start. Within two hours, we were able not only to convey the basic methodological concepts of GQM$^+$Strategies but also to initially *Characterize* the scope and environment for the GQM$^+$Strategies application. The environmental context can be described as safety-critical systems with a specific focus on software.

The scope for the application was to include JAXA's organizational top-level, the project level, as well as one specific organizational unit and potential external suppliers. The organizational unit involved in the application of GQM$^+$Strategies at JAXA drives continuous improvement in software development within the organization. One problem that may occur in the context of improvement initiatives is that improvement initiatives as such are not necessarily explicitly aligned towards top-level business goals or they struggle with showing their contribution to top-level business goals [8]. The organizational unit at JAXA was interested in highlighting its contribution and aligning its activities towards the top-level goals of the overall organization.

Potential suppliers were also part of the scope, as JAXA collaborates with many different suppliers. Improving the transparency of distributed projects through measurement was seen as a key component for the improvement of overall project performance and success. Particularly distributed collaborations provide challenges to measurement, as different understandings and motivations with respect to measurement have to be aligned.

In summary, the objectives consisted of explicitly modeling how the organizational unit involved in continuous improvement contributes to the top-level goals of JAXA. The other objective of the GQM$^+$Strategies application was to align and to make transparent the measurement needs throughout the organization in the context of distributed projects with external suppliers. One additional sub-condition that was defined for the GQM$^+$Strategies application was to reuse existing measurement assets.

Within the timeframe of the initial workshop, we were also able to elicit first context factors and assumptions as well as initial goals and strategies. Thus, we started the preparation for the main activities in *Set Goals*. This approach turned out to be useful, as it facilitated the preparation of the subsequent workshop.

**Iteration 1**

In preparation of this iteration and of the second joint workshop, an Internet research was performed in order to obtain another perspective on the organization's mission and vision, as well as on its top-level goals and strategies. The officially published material was a valuable complement to the internal view provided by JAXA members. Furthermore, measurement assets that could be reused were elicited. These included software standards and measures in use. Based on this initial information, we performed the gap analysis and created a first draft of the GQM$^+$Strategies including initial measurement goals.

We then conducted the second one-day workshop, in which the same manager and senior engineer as well as five additional engineers participated. As the group was larger than in the first workshop and more inexperienced persons participated, we started with a two-hour GQM$^+$Strategies tutorial and then continued with the session for defining the GQM$^+$Strategies grid. In the grid definition session, we started the discussion on the basis of the grid draft, beginning with the top-level goals of the organization. This initial draft, which was presented with the GQM$^+$Strategies visualization tool, enhanced understanding and was beneficial for our discussion. We were able to define the top-level of the grid and elicited further context information and assumptions during the discussion. After having defined the top-level of the grid, we turned to discussing the derivation and alignment of goals and strategies across all relevant organizational levels and units. The grid derivation process [5], existing templates, and tool support were helpful for formalizing goals and strategies and capturing context and assumptions in a systematic way. The focus of this session was clearly on achieving a joint view of the basic structure of goals, strategies, and the organization, as well as modeling the grid accordingly. For the discussion of different modeling alternatives, we additionally used a whiteboard, which facilitated the understanding of possible alternatives. Measurement was discussed on a very abstract level based on the initially defined measurement goals.

**Iteration 2**
Based on the joint understanding of top-level goals and strategies as well as the basic structure of the grid, further development was performed offline by JAXA and Fraunhofer engineers. Following the offline preparation, we had a joint video conference. In this video conference, the manager and senior engineer as well as another engineer tool part. It lasted approximately two hours. The focus of the video conference was on consolidating results and finalizing the GQM$^+$Strategies grid. Measurement was again only discussed on the level of measurement goals.

**Iteration 3**
After the second iteration, when the grid was close to completion, we started with the definition of the GQM measurement. Based on the defined measurement goals, the measurement models were refined by using the traditional GQM approach and refining measurement goals into questions and metrics. JAXA's existing standards and measurement samples provided the requirements with respect to the reuse of measurement assets. It was possible to integrate existing engineering measures into the GQM measurement models of the grid. But especially for top-level measures, we needed to define new GQM models and measures in order to be able to evaluate the success of the respective goals and strategies elements. The basic measurement definition was performed in preparation of the final video conference. In this video conference, the core group (manager, senior engineer, and engineer) took part. It lasted approximately two hours. The focus was on consolidating the results with respect to measurement and finalizing the GQM$^+$Strategies grid. Final adaptations that closed this iteration were performed offline.

**Results**
The modeling of the GQM$^+$Strategies grid and thus the definition of the strategic measurement system was the major objective of this collaboration. Figure 3 shows a representation of the resulting GQM$^+$Strategies grid structure. In total, the modeled grid contains 23 Goal$^+$Strategies elements and, consequently, 23 GQM measurement models were defined in order to measure and evaluate the success of the respective Goal$^+$Strategies elements.

Although this GQM$^+$Strategies application was divided into the four steps described with regard to defining and evolving the GQM$^+$Strategies grid, we did not track the effort accordingly. The total effort for this application of the approach amounted to approximately 18 person-days for Fraunhofer IESE and 9 person-days for JAXA (one person-day corresponds to eight hours).
The following section presents further details of the GQM$^+$Strategies grid.

## 4  The GQM$^+$Strategies Grid

The objectives of the GQM$^+$Strategies application were (1) to clarify and explicitly align the contribution of an internal organizational unit for software process improvement (JAXA JEDI/SPI) to the top-level goals of JAXA and (2) to align and make transparent the measurement needs throughout the organization in the context of collaborative projects with external suppliers. These objectives impose a multi-organizational setting for the modeling of the GQM⁺Strategies grid. Aspects of modeling as well as selected details of the resulting GQM⁺Strategies grid are discussed in the following.

### 4.1 Modeling the GQM⁺Strategies Grid

For both objectives, the alignment to top-level goals was relevant and therefore modeling of the top-level goals was necessary in order to guarantee a goal-oriented procedure. The representation of the multi-organizational setting with GQM⁺Strategies was achieved by modeling and integrating organizational constructs that represent the respective internal and external organizational units. Modeling these organizational constructs consisted of capturing their relevant organizational structures. Integration of these constructs was possible by linking them to an appropriate interface (organizational level) at JAXA. Then associated goals and strategies were refined within the structures of the respective organizational construct. At such an interface, many opportunities exist for interaction between the organizational units involved. For example: At an interface between JAXA and a supplier, it is possible not only to define goals or success criteria, but also to use the definition of measurement models to gain a higher level of insight into the actual implementation of the success criteria and to understand which strategies are pursued at the supplier organization to achieve success. Such opportunities provide insights that go beyond pure analysis of engineering measurement data.

For objective (1) the modeling and alignment of goals and strategies of an organizational unit were of importance. The organizational construct in question is the organizational hierarchy of the respective unit. Using this construct, we modeled the contribution of the software improvement unit and one exemplary contribution of suppliers. For both cases, the top-level of the organizational unit was aligned and linked to JAXA's top-level. The resulting structure resembled a line organization.

For objective (2), the measurement needs were transformed and formalized into goals at internal and external organizational levels. Measurement was seen as a means to achieve better project performance and hence the addressed organizational construct was JAXA's project organization. Requirements for the reuse of measurement assets were defined by internal standards and existing engineering measures. Goals and strategies were directly refined from JAXA's project level to the supplier's project level, representing the project organization at JAXA.

Thus, the GQM⁺Strategies grid does not only provide the possibility to integrate different organizational units but also captures aspects of the project and line organization of JAXA. The grid was refined from JAXA's organizational top-level into two, respectively three, additional organizational levels (see Figure 3). The project organization was modeled with the JAXA project-level goals being linked to JAXA's top-level and the supplier's project-level goals being linked to the JAXA project level (see Figure 3 on the right). Organizational aspects that capture line organizations use four levels. These include JAXA's top-level, the management level of each unit, a sub-unit level, and finally the operational level of each sub-unit (see Figure 3 on the left).

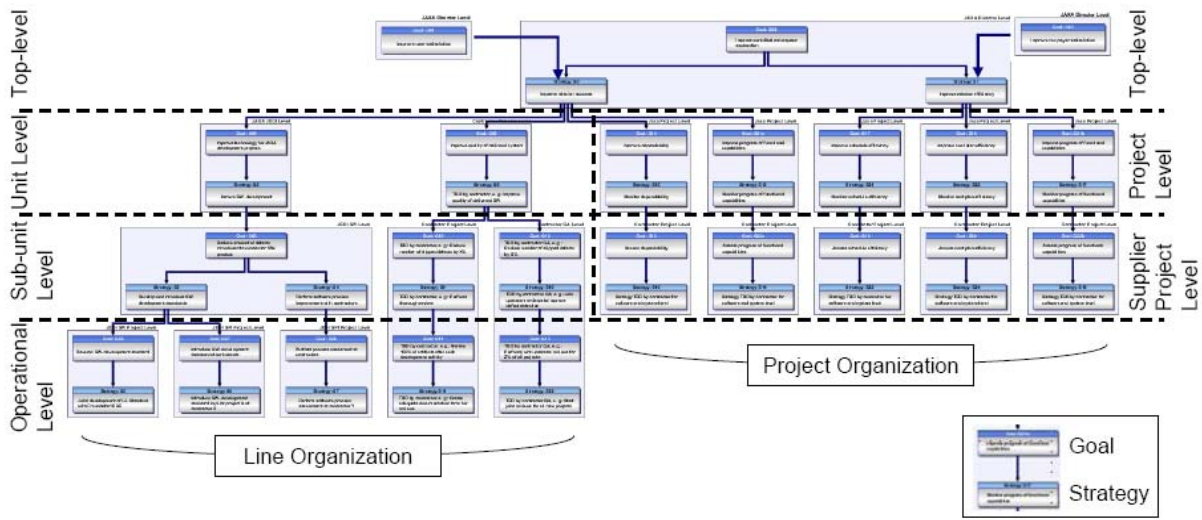

**Figure 3:** Structure of the GQM$^+$Strategies grid

This example shows that it is possible to model and align different organizational constructs in one GQM$^+$Strategies grid, including internal and external organizational units, and thus achieve alignment in a multi-organizational setting. After presenting these structural aspects of the grid, we continue with further details regarding its contents.

### 4.2 Selected Details from the GQM$^+$Strategies Grid

Gaining a deeper understanding of the GQM$^+$Strategies grid requires a discussion of the top-level goals and the different branches that refine the top-level. All branches link to the top-level goals and strategies that we found to be most important for JAXA. In the following, planned success magnitudes and time frames are omitted due to confidentiality reasons and the focus is placed on the general ideas. At the top-level, we find the goals "improve contribution to space exploration", "improve user satisfaction", and "improve tax payer satisfaction". The first goal is quite obvious for a space exploration agency. The realization of this goal is implemented by two strategies, "improve mission success" and "improve mission efficiency". Mission success can be linked directly to the space exploration goal. But mission success is not sufficient as available resources are limited. Missions should be performed efficiently, and thus evaluation of the actual contribution is a weighted combination of success and efficiency measures.

User satisfaction was linked to mission success, as the satisfaction of users is evaluated based on the achievement of quality and functional aspects of a system or service realized during a mission. A user in the sense of JAXA is anybody who uses JAXA systems or services.

Finally, tax-payer satisfaction was linked to mission efficiency based on the assumption that a sponsor does not want the financial resources he contributes to be wasted. Consequently, tax-payer satisfaction can be evaluated based on efficiency of resource usage. Additionally, a comparison between perceived value and financial resources that are provided can be performed, which, however, requires representative primary tax-payer data.

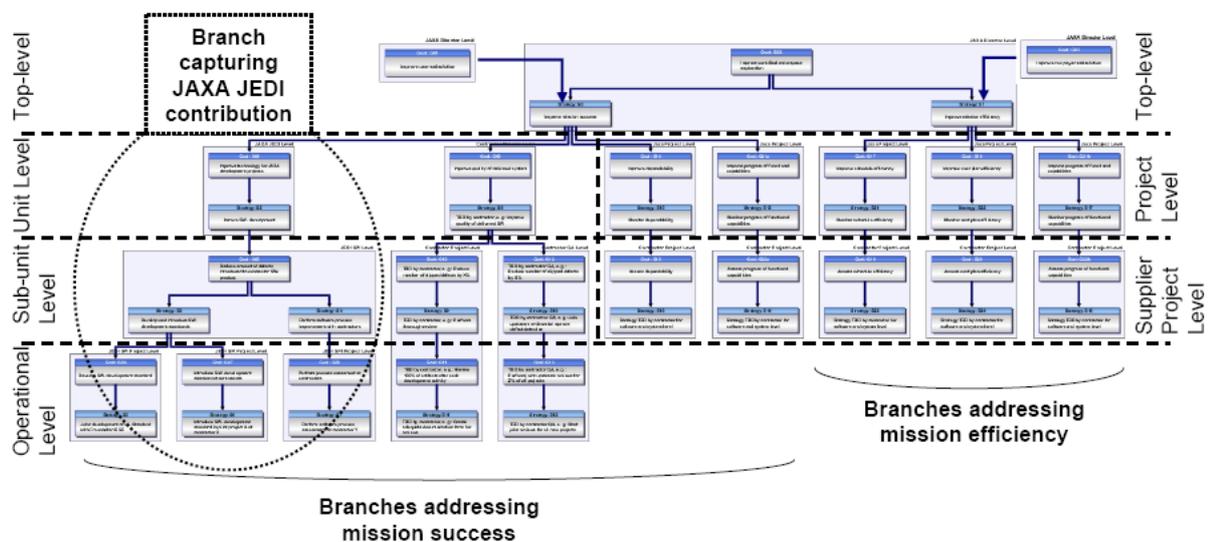

**Figure 4:** Contents addressed in the GQM⁺Strategies grid

The top-level goals are refined into branches and aligned to goals of internal and external organizational units. The three branches on the right side are linked and aligned to the "improve mission efficiency" strategy and deal in different ways with the improvement of project progress in the context of distributed JAXA projects. The basic assumption made here is that an increase in transparency in the context of distributed collaboration is the basis for efficiency improvement as project risks can be mitigated earlier and communication costs can be reduced. The three branches deal with the improvement of schedule efficiency, cost plan efficiency, as well as the progress of functional capabilities. These aspects are defined and made measurable at the JAXA project level and then refined to the supplier level.

The four branches on the left side are linked and aligned with the "improve mission success" strategy and deal with the improvement of the mission results with respect to quality. Two of these branches again represent the project structure and deal with improvement of dependability and realization of functional capabilities.

The other two address the direct contribution of suppliers and of the internal organizational unit to mission success.

In the following, we will discuss the branch that models the JAXA/JEDI SPI group contribution (see Figure 4) to provide a more detailed view on the GQM[+]Strategies grid. The JEDI SPI group deals with software process improvement and is a sub-unit of JAXA/JEDI. The purpose of JAXA/JEDI is to introduce new technologies within JAXA. JAXA/JEDI deals with many different technologies and application domains. Technological improvement is seen as one major contribution with respect to the improvement of mission success. Figure 5 shows the Goal[+]Strategies element at the JAXA/JEDI level (unit level in Figure 4). The Goal[+]Strategies element models the goal at the JAXA/JEDI level and refines a strategy that aims at achieving the goal. Software development "technologies" were relevant for our context. The template presented in Figure 5 was used for formalizing goals and strategies and captures associated context information and assumptions (see also [4]).

| Goal: | Improve technology for JAXA development projects | | | | | | |
|---|---|---|---|---|---|---|---|
| Activity | Focus | Object | Magnitude | Timeframe | Scope | Constraints | Relations |
| Improve | Technology | JAXA Project | X% over history | Until next status review | JAXA | Cost, schedule, people capabilities | (+) supplier level goals, JAXA project level goals |
| Context | ■ JEDI supports JAXA with new effective technologies | | | Assumptions | ■ Supporting the technological improvement within JAXA (by X%) helps to improve mission success by A% | | |
| Strategy: | Improve software (SW) development | | | | | | |
| Context | ■ Software development represents a critical aspect for JAXA missions | | | Assumptions | ■ Improving SW development capabilities in JAXA projects by B% will have a positive impact on JEDI mission success (C%) | | |

**Figure 5:** Goal[+]Strategies element

In order to evaluate the actual achievement of the goal, a GQM measurement model was defined that measures the technological improvement at the JEDI level (Figure 6). These measures are generic with respect to technologies and need to be refined for specific technologies, which, in our context, were software development technologies.

The JEDI SPI group (sub-unit level in Figure 4) is dealing with software development and in particular with the improvement of software development processes that are used in the context of JAXA projects. The main goals in contributing towards overall mission success is to reduce the number of defects that are introduced into software products developed for JAXA projects. The main strategies used by the SPI group to achieve its goal are the development of JAXA software development standards and the performance of software process improvement initiatives with their suppliers. Thus in order to evaluate success with respect to the JEDI SPI goal, an evaluation of the actual defect reduction is necessary for projects that use the "technology" disseminated by the JEDI SPI group. One level below, that is, on the operational-level of JEDI SPI, the upper-level strategies are further refined into operational level goals and strategies, namely concrete development and improvement initiatives performed by the JEDI SPI group.

| GQM Goal | Object | Purpose | Quality Focus | Viewpoint | Context |
|---|---|---|---|---|---|
| Technology improvement | Technology | Evaluation | Improvement | JAXA JEDI | JAXA Project |
| **Questions** | | | | | |
| ▪ Q1: What is the technological improvement provided by JEDI per application domain?<br>▪ Q1.1: What is the number of new technology introductions per application domain?<br>▪ Q1.2: What is the impact of an introduced technology?<br>　▫ Q1.2.1: What is the dissemination of the introduced technology?<br>　▫ Q1.2.2: What is the effectiveness of the introduced technology?<br>▪ Q2: What is the measurement baseline? | | | | | |
| **Metrics** | | | | | |
| Technology improvement | SUM(Impact) / number of technologies (per application domain) | | | | Q1 |
| Number of technologies | Number of technologies introduced per application domain | | | | Q1.1 |
| Impact | Dissemination*(Average effectiveness) | | | | Q1.2 |
| Dissemination | (Number of introductions of a specific technology) / (Number of possible introductions) | | | | Q1.2.1 |
| Effectiveness | Degree or ratio of improvement (e.g. defect reduction) | | | | Q1.2.2 |
| Measurement baselines | Measurement baselines for technology improvement | | | | Q2 |
| **Decision criteria** | | | | | |
| Technology improvement ≥ threshold ( measurement baseline or target) | | | | | |

**Figure 6:** GQM measurement model

## 5  Lessons Learned and Improvement Potentials

The application of the GQM⁺Strategies approach provided new insights into the existing concepts, from which we derived the following lessons learned (LL) and improvement potentials (IP).

LL1: The GQM⁺Strategies process was very helpful in structuring and performing the application of GQM⁺Strategies. Overall, the activities were performed in the prescribed manner, although the performed process was not completely sequential. In our application, there was a minor iteration between the activities Characterize and Set Goals. Furthermore, Set Goals was performed iteratively, with three main iterations.

IP1: Including possible iterations might enhance the GQM⁺Strategies process.

LL2: Besides eliciting existing goals, strategies, and measurement assets, the organizational structures played a major role for understanding the interactions within the organization and for the later integration and alignment.

IP2: A complementary formalization of organizational structures might prove to be a beneficial adaptation of the process. More emphasis could be placed on modeling the organizational setting, as the organizational structure has an influence on the structure of the GQM⁺Strategies grid.

LL3: With respect to the gap analysis it can be stated that it seemed to be most important to identify the assets with the highest relevance as well as the most significant misalignments and then to continue in a constructive manner with the integration and alignment of the different aspects. A gap analysis containing nearly all assets was not feasible in our case, as many documents needed to be translated, and thus we focused only on the most important.

IP3: Consequently, a cost-benefit evaluation or a scoping step for the gap analysis could be amended to the GQM⁺Strategies process.

LL4: So far, influences or relations between goals are captured and possible conflicts are identified. But the actual effects of these relations are addressed insufficiently.

IP4: What might be improved with respect to modeling is the concept of modeling interdependencies between different goals and strategies. For example, if a goal positively influences the outcome of another goal, the severity of this influence remains unknown. Modeling these aspects is crucial, especially considering the maintenance or evolution of such a grid. When goals and strategies start to change, it is important to know the effects of such changes on the remaining goals within the grid. In this scenario, it might also be necessary to capture and model external factors that influence goal achievement in a more systematic and formalized manner.

LL5: Finally, with respect to the evaluation of the results of the GQM⁺Strategies application, the approach proved to be capable of achieving the two stated objectives for the application, and our customers at JAXA were satisfied with the results.

IP5: This represents only a qualitative evaluation of the results, as we do not yet have a formal evaluation framework for GQM⁺Strategies.

## 6      Conclusions and Future Work

This paper presented first-hand experience from applying GQM⁺Strategies in a multi-organizational setting at the Japan Aerospace Exploration Agency (JAXA). We started with a short overview of the basic concepts of GQM⁺Strategies as well as the process of its application. We then discussed the process of applying GQM⁺Strategies at JAXA and showed how the GQM⁺Strategies process supports a structured implementation of the approach. The GQM⁺Strategies grid developed in the context of this collaboration helped to achieve the two major objectives. One objective of the GQM⁺Strategies application was to clarify and explicitly align the contribution of an internal organizational unit to top-level goals of JAXA. The other objective was to align and make transparent the measurement needs throughout the organization in the context of collaborative projects with external suppliers. We presented our approach of modeling the grid as well as its basic structure and selected details of modeled goals, strategies, and GQM measurement models. The existing approach and supporting tools and templates provide good support for modeling, and we were able to model a grid that fulfilled the expectations with respect to the project challenges and objectives.

Nevertheless, some improvement potential was identified concerning the process and the modeling. Future work could focus on enhancing the process with iterations and an additional cost-benefit analysis for scoping of some process activities better. Organizational structures could be embedded better into the approach, as they often have implications on the structure of the resulting grid. Furthermore, better handling of goal interdependencies and external environmental influence factors would be beneficial; in particular for the maintenance and evolution of GQM$^+$Strategies grids. Finally, a standardized evaluation framework for the evaluation of costs and benefits of the approach would be of great value for comparing future case studies. These aspects are on our research agenda.